\documentclass[useAMS,usenatbib]{mn2e}
\usepackage[dvips]{graphicx}
\newcommand{\bz}{$\langle B_z \rangle$}
\newcommand{\nz}{$\langle N_z \rangle$}
\newcommand{\vsini}{$v \sin i$}
\newcommand{\kms}{km\,s$^{-1}$}

\newcommand{\msun}{$M_\odot$}

\newcommand{\ra}{$R_{\rm A}$}

\title[$\epsilon$ Lupi: the first doubly-magnetic massive binary]{Detection of magnetic fields in both B-type components of the $\epsilon$ Lupi system: a new constraint on the origin of fossil fields?}
\author[M. Shultz]
{M. Shultz$^{1,2}$,
G.A. Wade$^{2}$,
E. Alecian$^{3,4,5}$
and the BinaMIcS Collaboration \\
$^1$Department of Physics, Engineering Physics \& Astronomy, Queen's University, Kingston, ON Canada, K7L 3N6 \\
$^2$Department of Physics, Royal Military College of Canada, Kingston, Ontario K7K 7B4, Canada\\
$^3$Universit\'e Grenoble Alpes, IPAG, F-38000 Grenoble, France\\
$^4$CNRS, IPAG, F-38000 Grenoble, France\\
$^5$LESIA, Observatoire de Paris, CNRS UMR 8109, UPMC, Universit\'e Paris Diderot, 5 place Jules Janssen, 92190, Meudon, France\\
}
\begin{document}

\date{}

\pagerange{\pageref{firstpage}--\pageref{lastpage}} \pubyear{2002}

\maketitle

\label{firstpage}

\begin{abstract}

High-resolution circular spectropolarimetric observations, obtained with ESPaDOnS in the context of the BinaMIcS Large Program, have revealed a magnetic field in the B3V secondary component of the SB2 binary system $\epsilon$ Lupi (B2/B3). As the B2V primary is already known to be magnetic, this is the first detection of a magnetic field in both components of an early-type binary system. The longitudinal magnetic field of the primary is $\sim -200$ G; that of the secondary $\sim +100$ G. Observations can be approximately reproduced by a model assuming the magnetic axes of the two stars are anti-aligned, and roughly parallel to their respective rotation axes. Estimated magnetospheric radii indicate a high probability that their magnetospheres are interacting. As many of the arguments for the different proposed formation scenarios of fossil magnetic fields rely upon evidence drawn from investigations of close binaries, in particular the rarity of magnetic ABO stars in close binaries and the previous absence of any known close binary with two magnetic, massive stars, this discovery may be an important new constraint on the origin of fossil magnetic fields. 

\end{abstract}


\section{Introduction}

A principal result of the Magnetism in Massive Stars (MiMeS) survey has been the demonstration that a significant subset ($\sim$10\%) of the early-type stellar population host strong (0.1-10 kG), stable magnetic fields (Wade et al., submitted). In contrast, magnetic fields are essentially ubiquitous amongst cool stars. The magnetic fields of magnetic, massive stars (MMSs) are currently explained as so-called fossil fields (e.g. \citealt{1945MNRAS.105..166C, 1977MNRAS.178...27M, 2004Natur.431..819B}): magnetic flux preserved from a previous era in a star's life.  Magnetohydrodynamic (MHD) simulations indicate that magnetic fields within a radiative envelope can relax into stable configurations which then dissipate due to Ohmic decay and other slow mechanisms on stellar evolutionary timescales \citep{1972MNRAS.156..419M, 2009MNRAS.397..763B}. This framework has been highly successful in explaining the observed characteristics of the magnetic fields of hot stars \citep{2015arXiv150200226N}, predicting magnetic topologies dominated by strong, global dipoles, with large angles between magnetic and rotational axes (e.g. \citealt{2010ApJ...724L..34D, 2011IAUS..272..178D, 2015IAUS..307..373E}). These characteristics are essentially identical to those long known for the magnetic Ap stars (e.g. \citealt{1980ApJS...42..421B}) and the He-weak and He-strong Bp stars (e.g. \citealt{1979ApJ...228..809B, 1983ApJS...53..151B}), suggesting a common mechanism behind the magnetic fields of all stars with radiative envelopes, i.e. from 1.5 to 50 \msun~\citep{1411.3604}. 

Notwithstanding these successes, the origin of fossil fields remains a subject of speculation, indicating a fundamental deficiency in our understanding of a basic physical property of all stars more massive than 1.5 \msun. One of the principal goals of the Binarity and Magnetic Interactions in various classes of Stars (BinaMIcS) project has been to investigate the origin of fossil fields amongst the MMSs \citep{2015IAUS..307..330A}, by examining in particular the properties and prevalence of binary systems containing MMSs. 


One of the BinaMIcS targets is $\epsilon$ Lupi (HD 136504, B2/B3), an SB2 system comprised of 2 early B-type stars with an orbital period of $\sim$4.6~d. Tentative evidence for a magnetic field in the system was presented by \cite{2009AN....330..317H}, based upon FORS1 data. The first reliable detection was confirmed with ESPaDOnS data by \cite{shultz2012}. Here we report for the first time the detection of a magnetic field in both components of $\epsilon$ Lupi, making it the first known close binary with two magnetic, massive stars.


\section{Observations}

Our dataset is comprised of 51 Canada-France-Hawaii Telescope (CFHT)/ESPaDOnS spectropolarimetric sequences. ESPaDOnS is a high-resolution ($\lambda/\Delta\lambda \sim 65,000$) spectropolarimeter with spectral coverage from 369.3 nm to 1048 nm across 40 overlapping spectral orders. Each sequence is comprised of 4 polarized subexposures, which are combined to yield unpolarized intensity (Stokes $I$), circularly polarized flux (Stokes $V$), and diagnostic null ($N$) spectra. On each night a minimum of 4, and a maximum of 11, sequences were obtained. Sequences obtained on the same night were co-added in order to increase the SNR, yielding 10 independent measurements. The first two were acquired by the MiMeS Large Program (LP), the most recent by the BinaMIcS LP, and the remainder in the context of PI program CFHT14AC010 (PI: M. Shultz). The log of observations is provided in Table \ref{bz_tab}. 

\begin{figure}
\centering
\includegraphics[width=8cm]{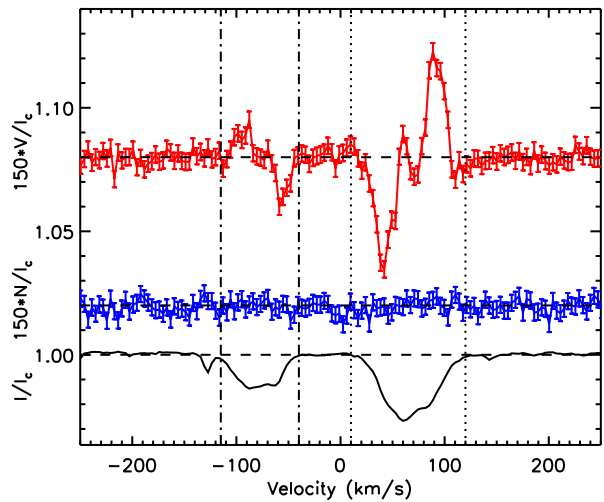}
\caption{LSD profile from the night of 2015/04/09. Dotted lines indicate the integration range for the primary, dot-dashed lines for secondary. Both stellar components are separated in Stokes $I$ (bottom); $N$ is consistent with noise (middle); in Stokes $V$, a definite detection is registered in both line profiles (top).}
\label{lsd}
\end{figure}

To further increase the SNR, we used the least-squares deconvolution procedure (LSD; \citealt{d1997}), in particular the iLSD package \citep{koch2010}. The line list used to construct the LSD mask was obtained from the Vienna Atomic Line Database VALD3 \citep{piskunov1995, ryabchikova1997, kupka1999, kupka2000}, using an `extract stellar' request. Since both components have similar temperatures, we used a single mask that we adjusted to best match the spectrum. We began with a 20~kK solar metallicity mask, cleaned to include only metallic lines unblended with H, He, or telluric lines, with 234 spectral lines remaining in the mask after cleaning. The remaining lines were `tweaked' by empirically adjusting the line depths to match the observed line strengths. LSD profiles were then extracted using 3.6 \kms~velocity pixels, with normalization values of the wavelength and Land\'e factor of $\lambda_0 = 500$ nm and $g_0 = 1.2$. Noise was suppressed using a Tikhonov regularization factor of 0.2 \citep{koch2010}. The LSD profile extracted from the BinaMIcS observation (the highest SNR measurement, obtained on HJD 2457122) is shown in Fig. \ref{lsd}. 


To quantify the presence of a magnetic signature, we applied two diagnostic techniques to the LSD profiles. The first was to calculate the False Alarm Probability (FAP) of the signal in Stokes $V$ inside the line profile \citep{d1997}. Detections are considered definite (DD) if FAP~$< 10^{-5}$, and non-detections (ND) if FAP~$> 10^{-3}$ \citep{d1997}. Integration ranges were set individually for each LSD profile, depending on the radial velocities and line-profile widths of the components, as illustrated in Fig. \ref{lsd}. Detection flags are provided in Table \ref{bz_tab}. The most recent, high-SNR measurement yields a DD in the secondary component.




\begin{figure}
\centering
\includegraphics[width=8cm]{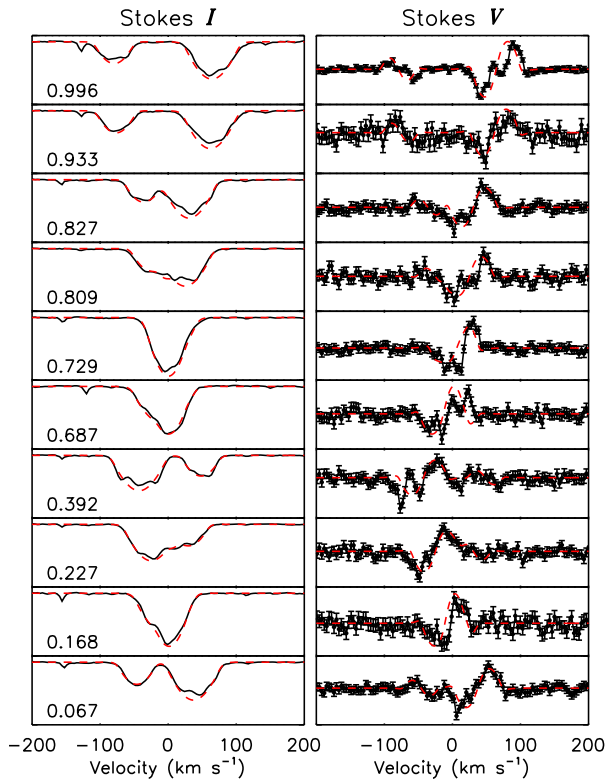}
\caption{Observed (solid black lines) and synthetic (dashed red lines) LSD profiles arranged in order of orbital phase. The BinaMIcS observation is at phase 0.996.}
\label{lsd_binphase}
\end{figure}

\begin{figure}
\centering
\includegraphics[width=8cm]{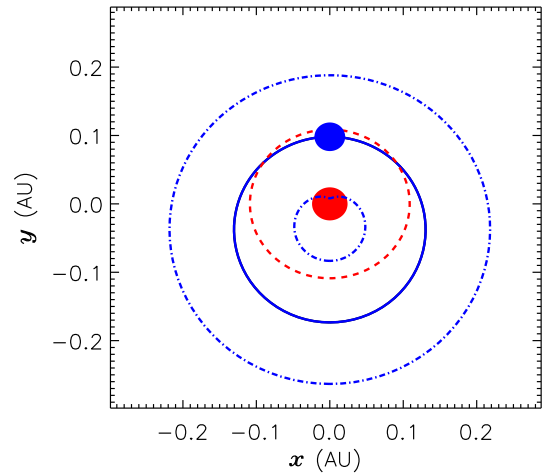}
\caption{Schematic of the $\epsilon$ Lupi system's orbit, in the inertial frame of the primary. The primary is indicated by the solid (red) circle in the centre, the secondary by the solid (blue) circle at periastron. The orbit of the secondary is indicated by the solid (blue) line. The dashed (red) line indicates the primary's \ra. The dot-dashed (blue) lines indicate the inner and outer extent of the secondary's \ra~during an orbital cycle. Orbits and stellar radii are shown to scale.}
\label{orbit}
\end{figure}

\begin{table*}
\centering
\caption[Observations]{Log of observations and magnetic measurements. SNR is per 1.8 \kms~spectral pixel. DF refers to the detection flag, as discussed in the text. Columns 7--9 (`Combined') provide magnetic measurements performed including both stellar line profiles; at phases during which the profiles are blended, ony combined measurements are given. Remaining columns give magnetic measurements for the individual stellar components, at phases at which the components' line profiles are well-separated.}
\resizebox{18 cm}{!}{
\begin{tabular}{rcrrrrrrrrrrr}
\hline
\hline
 & & & & \multicolumn{3}{c}{Combined} & \multicolumn{3}{c}{Primary} & \multicolumn{3}{c}{Secondary} \\
HJD - & UT Date & $\phi_{\rm orb}$ & SNR & \bz & \nz & DF & \bz & \nz & DF & \bz & \nz & DF \\
2450000 & &  & & (G) & (G) &  &  (G) & (G) &  &  (G) & (G) &  \\
\\
\hline
\\
7122.0250 & 2015-04-09 & 0.996 & 4904 & -102$\pm$30 &  -43$\pm$30 & DD & -183$\pm$12 & -16$\pm$12 & DD & 104$\pm$14 &  13$\pm$14 & DD \\
6824.8751 & 2014-06-16 & 0.827 & 1642 & -127$\pm$42 &    4$\pm$42 & DD & -240$\pm$25 & -28$\pm$25 & DD &  78$\pm$29 &   8$\pm$29 & ND \\
6822.8914 & 2014-06-14 & 0.392 & 1343 & -119$\pm$60 &  -39$\pm$60 & DD & -228$\pm$37 & -59$\pm$37 & DD & 104$\pm$33 & -51$\pm$33 & ND \\
6821.8707 & 2014-06-13 & 0.168 & 1212 & -102$\pm$28 &   -7$\pm$28 & DD & -- & -- & -- & --  & -- & -- \\
6819.8677 & 2014-06-11 & 0.729 & 2239 & -115$\pm$13 &   19$\pm$13 & DD & -- & -- & -- & --  & -- & -- \\
6816.8524 & 2014-06-08 & 0.067 & 1779 & -196$\pm$45 &  -65$\pm$45 & DD & -220$\pm$24 & -16$\pm$24 & DD &  83$\pm$29 &   2$\pm$29 & ND \\
6760.9572 & 2014-04-13 & 0.809 & 1326 & -103$\pm$37 &   -8$\pm$37 & DD & -- & -- & -- & --  & -- & -- \\
6756.9622 & 2014-04-09 & 0.933 & 1261 &  -94$\pm$95 & -100$\pm$95 & DD & -175$\pm$44 & -33$\pm$44 & DD & 147$\pm$46 & -57$\pm$46 & ND \\
5727.8117 & 2011-06-15 & 0.227 & 1461 & -127$\pm$33 &   19$\pm$33 & DD & -- & -- & -- & --  & -- & -- \\
5634.1538 & 2011-03-13 & 0.687 & 1325 & -124$\pm$22 &  -17$\pm$22 & DD & -- & -- & -- & --  & -- & -- \\
\hline
\hline
\end{tabular}
}
\label{bz_tab}
\end{table*}

The longitudinal magnetic field \bz~was measured from the LSD profiles by taking the first-order moment of the Stokes $V$ profile \citep{mat1989}. The same integration ranges were used for measuring \bz~as for evaluation of the FAPs. \bz~values for combined line profiles are provided at all orbital phases in Table \ref{bz_tab}, and for the primary and secondary components individually at phases at which the components' lines are separated. We also measured \nz~using the $N$ profiles: it is consistent with noise in both the blended profiles and the individual profiles at all phases. 

Prior to the clear detection of the magnetic signature of the secondary in April 2015, analysis of data collected between 2011 and 2014 suggested that the secondary component might also possess a magnetic field. At phases in which the components are clearly separated, the secondary consistently registers \bz~$\sim +100$ G, while the magnetic field of the primary is always $\sim - 200$ G. When the components are blended, \bz~$\sim - 100$ G, as might be expected if two Zeeman signature of opposite polarity are partially cancelling. All 7 FORS1 measurements are also $\sim -100$ G \citep{hub2011a}: due to the low spectral resolution of this instrument, the components are likely blended at all orbital phases.

\section{Discussion}


$\epsilon$ Lupi's orbital elements were most recently determined by \cite{uytterhoeven2005} (hereafter U05): $e=0.277$, $\omega = 18^\circ$, $P_{\rm orb}\sim4.56~{\rm d}, i = 21^\circ$. The LSD profiles are shown phased with the U05 ephemeris in Fig. \ref{lsd_binphase}. U05 found \vsini$~= 42$~\kms~for the primary and 37~\kms~for the secondary. In the left panels of Fig. \ref{lsd_binphase} we show model Stokes $I$ LSD profiles. The basic disk integration model used in synthesizing the LSD profiles is described by \cite{petit2012a}. The model approximates the LSD profile as an individual spectral line, rather than extracting an LSD profile from a synthetic spectrum; while there are limitations to this approach \citep{koch2010}, it is sufficient for a first-order model of the observed LSD profiles, and much cheaper computationally. Due to the inclusion of an {\em ad hoc} turbulent broadening, chosen to optimize the fit to the observed profiles \citep{petit2012a}, we use slightly lower \vsini~values than those given by U05: \vsini$_{\rm P} = 37$ \kms and \vsini$_{\rm S} = 27$ \kms. The synthetic LSD profiles corresponding to the individual stars were centred at radial velocities computed using the U05 orbital elements, and then added using a flux ratio of 70\%/30\% primary/secondary, matching the observed line strength ratio. The combined profiles were then normalized to reproduce the equivalent widths of the observed LSD profiles. 


The right panels of Fig. \ref{lsd_binphase} show a comparison between observed and synthetic Stokes $V$ profiles. While the rotational periods, inclinations $i$ of the rotational axes, obliquity angles $\beta$ between the rotational and magnetic axes, and dipolar field strengths $B_{\rm d}$ cannot yet be determined with certainty, some inferences can be made from the existing dataset. At phases during which the profiles are well-separated, \bz~is constant within the error bars, while at blended phases, \bz~is always about $-100$ G. This suggests that \bz~of both components is roughly constant, implying that either $i$, $\beta$, or both, are relatively small (considering the known \vsini s, the other possibility - that the periods are much longer than the timescale over which the data were acquired - is not considered). We therefore set $i$ to the orbital inclination of 21$^\circ$ for both stars, and fixed $\beta = 0^\circ$ such that all variation in Stokes $V$ is due to orbital modulation, consisting only of RV shifts applied to constant Stokes $V$ profiles. We adopted $B_{\rm d, P} = 900$ G and $B_{\rm d, S} = 600$ G, as being approximately the minimum strengths compatible with the observed \bz.

As is clear from Fig. \ref{lsd_binphase}, this model reproduces the gross features of Stokes $I$ and $V$ at all phases, including those during which the line profiles are blended. In many cases, small-scale features are present in the observed Stokes $V$ profiles. These are likely a result of non-radial pulsation (U05, \citealt{2006AA...452..945T}), as indicated by comparisons between the observed Stokes $I$ and $V$ profiles: strong asymmetries in $I$ are typically accompanied by corresponding asymmetries in $V$, as is seen most clearly at phase 0.392. 


Determining the rotational periods and magnetic properties of these stars will require a larger, high-SNR spectropolarimetric dataset. Modelling the nonradial pulsations will be necessary, as these effects may well dominate the variability of Stokes $V$ for each star; as these frequencies and modes are unidentified (U05), large, high-quality photometric and spectroscopic datasets will also be needed. 


We evaluate the strength of magnetic wind confinement via the magnetic wind confinement parameter $\eta_*$ (Eqn. 7 of \citealt{ud2002}), using the mass-loss recipe of \cite{vink2001}. Taking $R_{\rm *,P} = 4.9 \pm 0.8 R_\odot$, $M_{\rm *,P} = 8.7 \pm 1.0 M_\odot$, $\log{T_{\rm eff,P}} = 4.34 \pm 0.02$, $R_{\rm *,S} = 4.0 \pm 0.5 R_\odot$, $M_{\rm *, S} = 7.3 \pm 0.9 M_\odot$, and $\log{T_{\rm eff,S}} = 4.31 \pm 0.03$ (U05), and $B_{\rm d, P} = 0.82$ kG and $B_{\rm d, S} = 0.5$ kG (the minimum values compatible with \bz), we find $\eta_* > 250$ for both stars. Since $\eta_* >> 1$, their winds are magnetically confined above the stellar surface. The physical extent of their magnetospheres, given by the Alfv\'en radius \ra~(Eqn. 10, \citealt{ud2008}), is $R_{\rm A,P} > 4.47 R_*$ and $R_{\rm A,S} > 4.26 R_*$. 

In a close binary system it is of interest to compare \ra~to the physical separation between the stars. U05 found semi-minor and semi-major axes of $a_1\sin{i}\sim 0.022$ AU and $a_2\sin{i} \sim 0.026$ AU, or about 0.061 AU and 0.073 AU in the orbital plane, assuming an orbital inclination of 21$^\circ$. This is only about 3 $R_{\rm *,P}$ or 4 $R_{\rm *,S}$. Fig. \ref{orbit} shows a schematic of the orbit, in the rest frame of the primary, overlaid with the Alfv\'en radii \ra~of both stars. We use Alfv\'en radii calculated upon the assumption of minimum $B_{\rm d}$ (rather than the slightly higher values adopted in Fig. \ref{lsd_binphase}). There is substantial overlap between the two magnetic confinement regions at all points during the orbit. Since Fig. \ref{orbit} involves lower bounds for $B_{\rm d}$ and \ra, this suggests that their magnetospheres are very likely to be continuously interacting. If the stars' rotational periods are synchronized, magnetically-induced spindown \citep{wd1967, ud2009} may well be translated into loss of orbital angular momentum and consequent shrinking of the orbit \citep{2009MNRAS.395.2268B}. If the stars are not rotating synchronously, there may be magnetic reconnection phenomena due to the relative motion of the magnetospheres (e.g. \citealt{2014IAUS..302...44G}). Three-dimensional magnetohydrodynamic simulations (e.g. \citealt{ud2013}) will be necessary in order to evaluate the importance of such effects. 

There is no evidence of emission in optical or ultraviolet lines \citep{petit2013}. The absence of H$\alpha$ emission is expected, given the weak magnetic fields and winds, and the probable slow rotation as inferred from the low values of \vsini. The magnetic B2 star HD 3360, which has similar magnetospheric properties to those of the $\epsilon$ Lupi components, shows periodic UV variability \citep{oskinova2011, petit2013}. However, as there is only 1 high-dispersion IUE spectrum of $\epsilon$ Lupi, UV variability cannot be evaluated. The system is overluminous in X-rays, as expected for a magnetic massive star, and indeed is precisely on the X-ray luminosity trend predicted by the semi-analytic XADM model \citep{ud2014, 2014ApJS..215...10N}, albeit somewhat harder than other magnetic, massive stars with similar properties \citep{2014ApJS..215...10N}. 

Since the separation between the two stars is relatively small, it may be reasonable to consider that the magnetic field detected in the secondary is induced by the field of the primary. However, the relatively weak magnetic field of the primary argues against this: at the distance of the secondary, the maximum magnetic field strength of the primary's dipole field is only a few G, assuming a $1/r^3$ decay. This is an order of magnitude weaker than the measured \bz~of the secondary. In order to produce \bz$\sim$100 G at the distance of the secondary, the primary's magnetic dipole would need to be $\sim$15 kG, which is clearly incompatible with the data. Moreover, it is unclear how an induced field would explain the apparently anti-aligned fields of the two stars.

The discovery of a doubly-magnetic early-type close binary system may generate new insights into our understanding of the origin of fossil magnetic fields. The MiMeS survey showed MMSs to be relatively rare, approximately 10\% of the population of bright OB stars in the Galaxy \citep{grun2012c}. In remarkable contrast, out of 151 close binary systems observed in the context of the BinaMIcS survey, not one  new MMS has been detected, and in the broader sample only one magnetic field (in an F5 star) has been detected. This implies an incidence of magnetic stars in close binary systems of no more than $\sim$2\% \citep{2015IAUS..307..330A}. 

Two scenarios have been proposed for the origins of fossil magnetic fields: binary mergers \citep{2009MNRAS.400L..71F}, and magnetic flux preserved from the star formation process and amplified during early pre-main sequence evolution \citep{1999stma.book.....M}. 

The merger scenario proposes that the magnetic flux is generated by powerful dynamos excited during the merging event. The fraction of massive stars expected to undergo mergers while on the main sequence is roughly similar to the observed fraction of magnetic stars \citep{2012Sci...337..444S}. This scenario seems unlikely to account for the existence of a doubly-magnetic massive binary. Two separate mergers would be required, but the loss of orbital angular momentum required for one binary pair to merge is likely to be transferred to one or both of the other stars, thus widening or even destroying the orbit. 

In the second scenario, strong magnetic fields threading molecular clouds provide seed fields which are then amplified during the pre-main sequence convective phase. In this case, the historical failures to detect a doubly-magnetic hot star binary are a statistical artifact arising from the inherent rarity of MMSs: only $\sim$10\% of massive stars are magnetic, so only $\sim$1\% of close massive binaries should contain two magnetic stars. This scenario is supported by the similar incidence of magnetic fields, and the similar magnetic properties, of magnetic HeAe/Be stars and main sequence MMSs, indicating that the presence of a fossil field has already been established during the pre-main sequence \citep{2013MNRAS.429.1001A}. MHD simulations have also shown that strong magnetic fields inhibit fragmentation of molecular clouds \citep{2007MNRAS.377...77P, 2011ApJ...742L...9C}. Since close binaries are thought to be primordial, rather than the result of gravitational capture \citep{1994MNRAS.271..999B}, the rarity of close magnetic binaries may also be explained by this inhibition. The discovery of a close binary with two MMSs might be thought to be inconsistent with magnetic inhibition of fragmentation; however, the relatively weak magnetic fields of the $\epsilon$ Lupi system may be compatible with an origin in a weakly magnetized core, within which fragmentation may still occur \citep{2011ApJ...742L...9C}. However, this scenario is not without difficulties, as simulations of magnetized core collapse have tended to show that magnetic flux is not preserved (Masson et al., priv. comm.)  
\\\\
{\small {\em Acknowledgements} This work has made use of the VALD database, operated at Uppsala University, the Institute of Astronomy RAS in Moscow, and the University of Vienna. This work is based on observations obtained at the Canada-France-Hawaii Telescope (CFHT) which is operated by the National Research Council of Canada, the Institut National des Sciences de l'Univers of the Centre National de la Recherche Scientifique of France, and the University of Hawaii. All authors acknowledge the advice and assistance provided on this and related projects by the members of the BinaMIcS and MiMeS collaborations. M.S. acknowledges Jason Grunhut, whose binary LSD profile fitting software proved useful in the initial analysis, and Thomas Rivinius, who provided useful advice on observing strategies. G.A.W. acknowledges Discovery Grant support from the Natural Sciences and Engineering Research Council. E.A. acknowledges financial support from the Programme National de Physique Stellaire (PNPS) of INSU/CNRS.}

\bibliography{bib_dat.bib}{}

\begin{thebibliography}{47}
\expandafter\ifx\csname natexlab\endcsname\relax\def\natexlab#1{#1}\fi

\bibitem[{{Alecian} {et~al}\mbox{.}(2015){Alecian}, {Neiner}, {Wade}, {Mathis},
  {Bohlender}, {C{\'e}bron}, {Folsom}, {Grunhut}, {Le Bouquin}, {Petit},
  {Sana}, {Tkachenko}, \& {ud-Doula}}]{2015IAUS..307..330A}
{Alecian} E. {et~al.}, 2015, in IAU Symposium, Vol. 307, IAU Symposium, pp.
  330--335

\bibitem[{{Alecian} {et~al}\mbox{.}(2013){Alecian}, {Wade}, {Catala},
  {Grunhut}, {Landstreet}, {Bagnulo}, {B{\"o}hm}, {Folsom}, {Marsden}, \&
  {Waite}}]{2013MNRAS.429.1001A}
{Alecian} E. {et~al.}, 2013, \mnras, 429, 1001

\bibitem[{{Barker} \& {Ogilvie}(2009)}]{2009MNRAS.395.2268B}
{Barker} A.~J., {Ogilvie} G.~I., 2009, \mnras, 395, 2268

\bibitem[{{Bonnell} \& {Bate}(1994)}]{1994MNRAS.271..999B}
{Bonnell} I.~A., {Bate} M.~R., 1994, \mnras, 271, 999

\bibitem[{{Borra} \& {Landstreet}(1979)}]{1979ApJ...228..809B}
{Borra} E.~F., {Landstreet} J.~D., 1979, \apj, 228, 809

\bibitem[{{Borra} \& {Landstreet}(1980)}]{1980ApJS...42..421B}
{Borra} E.~F., {Landstreet} J.~D., 1980, \apjs, 42, 421

\bibitem[{{Borra} {et~al}\mbox{.}(1983){Borra}, {Landstreet}, \&
  {Thompson}}]{1983ApJS...53..151B}
{Borra} E.~F., {Landstreet} J.~D., {Thompson} I., 1983, \apjs, 53, 151

\bibitem[{{Braithwaite}(2009)}]{2009MNRAS.397..763B}
{Braithwaite} J., 2009, \mnras, 397, 763

\bibitem[{{Braithwaite} \& {Spruit}(2004)}]{2004Natur.431..819B}
{Braithwaite} J., {Spruit} H.~C., 2004, \nat, 431, 819

\bibitem[{{Commer{\c c}on} {et~al}\mbox{.}(2011){Commer{\c c}on}, {Hennebelle},
  \& {Henning}}]{2011ApJ...742L...9C}
{Commer{\c c}on} B., {Hennebelle} P., {Henning} T., 2011, \apjl, 742, L9

\bibitem[{{Cowling}(1945)}]{1945MNRAS.105..166C}
{Cowling} T.~G., 1945, \mnras, 105, 166

\bibitem[{{Donati} {et~al}\mbox{.}(1997){Donati}, {Semel}, {Carter}, {Rees}, \&
  {Collier Cameron}}]{d1997}
{Donati} J.-F., {Semel} M., {Carter} B.~D., {Rees} D.~E., {Collier Cameron} A.,
  1997, MNRAS, 291, 658

\bibitem[{{Duez} {et~al}\mbox{.}(2010){Duez}, {Braithwaite}, \&
  {Mathis}}]{2010ApJ...724L..34D}
{Duez} V., {Braithwaite} J., {Mathis} S., 2010, \apjl, 724, L34

\bibitem[{{Duez} {et~al}\mbox{.}(2011){Duez}, {Braithwaite}, \&
  {Mathis}}]{2011IAUS..272..178D}
{Duez} V., {Braithwaite} J., {Mathis} S., 2011, in IAU Symposium, Vol. 272, IAU
  Symposium, {Neiner} C., {Wade} G., {Meynet} G., {Peters} G., eds., pp.
  178--179

\bibitem[{{Emeriau} \& {Mathis}(2015)}]{2015IAUS..307..373E}
{Emeriau} C., {Mathis} S., 2015, in IAU Symposium, Vol. 307, IAU Symposium, pp.
  373--374

\bibitem[{{Ferrario} {et~al}\mbox{.}(2009){Ferrario}, {Pringle}, {Tout}, \&
  {Wickramasinghe}}]{2009MNRAS.400L..71F}
{Ferrario} L., {Pringle} J.~E., {Tout} C.~A., {Wickramasinghe} D.~T., 2009,
  \mnras, 400, L71

\bibitem[{{Gregory} {et~al}\mbox{.}(2014){Gregory}, {Holzwarth}, {Donati},
  {Hussain}, {Montmerle}, {Alecian}, {Alencar}, {Argiroffi}, {Audard},
  {Bouvier}, {Damiani}, {G{\"u}del}, {Huenemoerder}, {Kastner}, {Maggio},
  {Sacco}, \& {Wade}}]{2014IAUS..302...44G}
{Gregory} S.~G. {et~al.}, 2014, in IAU Symposium, Vol. 302, IAU Symposium, pp.
  44--45

\bibitem[{{Grunhut} {et~al}\mbox{.}(2012){Grunhut}, {Wade}, \& {MiMeS
  Collaboration}}]{grun2012c}
{Grunhut} J.~H., {Wade} G.~A., {MiMeS Collaboration}, 2012, in Astronomical
  Society of the Pacific Conference Series, Vol. 465, Proceedings of a
  Scientific Meeting in Honor of Anthony F. J. Moffat, {Drissen} L., {Rubert}
  C., {St-Louis} N., {Moffat} A.~F.~J., eds., p.~42

\bibitem[{{Hubrig} {et~al}\mbox{.}(2009){Hubrig}, {Briquet}, {De Cat},
  {Sch{\"o}ller}, {Morel}, \& {Ilyin}}]{2009AN....330..317H}
{Hubrig} S., {Briquet} M., {De Cat} P., {Sch{\"o}ller} M., {Morel} T., {Ilyin}
  I., 2009, Astronomische Nachrichten, 330, 317

\bibitem[{{Hubrig} {et~al}\mbox{.}(2011){Hubrig}, {Ilyin}, {Sch{\"o}ller},
  {Briquet}, {Morel}, \& {De Cat}}]{hub2011a}
{Hubrig} S., {Ilyin} I., {Sch{\"o}ller} M., {Briquet} M., {Morel} T., {De Cat}
  P., 2011, \apjl, 726, L5

\bibitem[{{Kochukhov} {et~al}\mbox{.}(2010){Kochukhov}, {Makaganiuk}, \&
  {Piskunov}}]{koch2010}
{Kochukhov} O., {Makaganiuk} V., {Piskunov} N., 2010, \aap, 524, A5

\bibitem[{{Kupka} {et~al}\mbox{.}(1999){Kupka}, {Piskunov}, {Ryabchikova},
  {Stempels}, \& {Weiss}}]{kupka1999}
{Kupka} F.~G., {Piskunov} N., {Ryabchikova} T.~A., {Stempels} H.~C., {Weiss}
  W.~W., 1999, \aaps, 138, 119

\bibitem[{{Kupka} {et~al}\mbox{.}(2000){Kupka}, {Ryabchikova}, {Piskunov},
  {Stempels}, \& {Weiss}}]{kupka2000}
{Kupka} F.~G., {Ryabchikova} T.~A., {Piskunov} N.~E., {Stempels} H.~C., {Weiss}
  W.~W., 2000, Baltic Astronomy, 9, 590

\bibitem[{{Mathys}(1989)}]{mat1989}
{Mathys} G., 1989, FCPh, 13, 143

\bibitem[{{Mestel}(1999)}]{1999stma.book.....M}
{Mestel} L., 1999, {Stellar magnetism}

\bibitem[{{Mestel} \& {Moss}(1977)}]{1977MNRAS.178...27M}
{Mestel} L., {Moss} D.~L., 1977, \mnras, 178, 27

\bibitem[{{Mestel} \& {Takhar}(1972)}]{1972MNRAS.156..419M}
{Mestel} L., {Takhar} H.~S., 1972, \mnras, 156, 419

\bibitem[{{Naz{\'e}} {et~al}\mbox{.}(2014){Naz{\'e}}, {Petit}, {Rinbrand},
  {Cohen}, {Owocki}, {ud-Doula}, \& {Wade}}]{2014ApJS..215...10N}
{Naz{\'e}} Y., {Petit} V., {Rinbrand} M., {Cohen} D., {Owocki} S., {ud-Doula}
  A., {Wade} G.~A., 2014, \apjs, 215, 10

\bibitem[{{Neiner} {et~al}\mbox{.}(2015){Neiner}, {Mathis}, {Alecian},
  {Emeriau}, {Grunhut}, {BinaMIcS}, \& {MiMeS
  collaborations}}]{2015arXiv150200226N}
{Neiner} C., {Mathis} S., {Alecian} E., {Emeriau} C., {Grunhut} J., {BinaMIcS}
  t., {MiMeS collaborations}, 2015, ArXiv:1502.00226

\bibitem[{{Oskinova} {et~al}\mbox{.}(2011){Oskinova}, {Todt}, {Ignace},
  {Brown}, {Cassinelli}, \& {Hamann}}]{oskinova2011}
{Oskinova} L.~M., {Todt} H., {Ignace} R., {Brown} J.~C., {Cassinelli} J.~P.,
  {Hamann} W.-R., 2011, \mnras, 416, 1456

\bibitem[{{Petit} {et~al}\mbox{.}(2013){Petit}, {Owocki}, {Wade}, {Cohen},
  {Sundqvist}, {Gagn{\'e}}, {Ma{\'{\i}}z Apell{\'a}niz}, {Oksala}, {Bohlender},
  {Rivinius}, {Henrichs}, {Alecian}, {Townsend}, {ud-Doula}, \& {MiMeS
  Collaboration}}]{petit2013}
{Petit} V. {et~al.}, 2013, \mnras, 429, 398

\bibitem[{{Petit} \& {Wade}(2012)}]{petit2012a}
{Petit} V., {Wade} G.~A., 2012, MNRAS, 420, 773

\bibitem[{{Piskunov} {et~al}\mbox{.}(1995){Piskunov}, {Kupka}, {Ryabchikova},
  {Weiss}, \& {Jeffery}}]{piskunov1995}
{Piskunov} N.~E., {Kupka} F., {Ryabchikova} T.~A., {Weiss} W.~W., {Jeffery}
  C.~S., 1995, \aaps, 112, 525

\bibitem[{{Price} \& {Bate}(2007)}]{2007MNRAS.377...77P}
{Price} D.~J., {Bate} M.~R., 2007, \mnras, 377, 77

\bibitem[{{Ryabchikova} {et~al}\mbox{.}(1997){Ryabchikova}, {Piskunov},
  {Kupka}, \& {Weiss}}]{ryabchikova1997}
{Ryabchikova} T.~A., {Piskunov} N.~E., {Kupka} F., {Weiss} W.~W., 1997, Baltic
  Astronomy, 6, 244

\bibitem[{{Sana} {et~al}\mbox{.}(2012){Sana}, {de Mink}, {de Koter}, {Langer},
  {Evans}, {Gieles}, {Gosset}, {Izzard}, {Le Bouquin}, \&
  {Schneider}}]{2012Sci...337..444S}
{Sana} H. {et~al.}, 2012, Science, 337, 444

\bibitem[{{Shultz} {et~al}\mbox{.}(2012){Shultz}, {Wade}, {Grunhut}, {Bagnulo},
  {Landstreet}, {Neiner}, {Alecian}, {Hanes}, \& {MiMeS
  Collaboration}}]{shultz2012}
{Shultz} M. {et~al.}, 2012, \apj, 750, 2

\bibitem[{{Telting} {et~al}\mbox{.}(2006){Telting}, {Schrijvers}, {Ilyin},
  {Uytterhoeven}, {De Ridder}, {Aerts}, \& {Henrichs}}]{2006AA...452..945T}
{Telting} J.~H., {Schrijvers} C., {Ilyin} I.~V., {Uytterhoeven} K., {De Ridder}
  J., {Aerts} C., {Henrichs} H.~F., 2006, \aap, 452, 945

\bibitem[{{ud-Doula} {et~al}\mbox{.}(2014){ud-Doula}, {Owocki}, {Townsend},
  {Petit}, \& {Cohen}}]{ud2014}
{ud-Doula} A., {Owocki} S., {Townsend} R., {Petit} V., {Cohen} D., 2014,
  \mnras, 441, 3600

\bibitem[{{ud-Doula} \& {Owocki}(2002)}]{ud2002}
{ud-Doula} A., {Owocki} S.~P., 2002, ApJ, 576, 413

\bibitem[{{ud-Doula} {et~al}\mbox{.}(2008){ud-Doula}, {Owocki}, \&
  {Townsend}}]{ud2008}
{ud-Doula} A., {Owocki} S.~P., {Townsend} R.~H.~D., 2008, MNRAS, 385, 97

\bibitem[{{ud-Doula} {et~al}\mbox{.}(2009){ud-Doula}, {Owocki}, \&
  {Townsend}}]{ud2009}
{ud-Doula} A., {Owocki} S.~P., {Townsend} R.~H.~D., 2009, MNRAS, 392, 1022

\bibitem[{{ud-Doula} {et~al}\mbox{.}(2013){ud-Doula}, {Sundqvist}, {Owocki},
  {Petit}, \& {Townsend}}]{ud2013}
{ud-Doula} A., {Sundqvist} J.~O., {Owocki} S.~P., {Petit} V., {Townsend}
  R.~H.~D., 2013, \mnras, 428, 2723

\bibitem[{{Uytterhoeven} {et~al}\mbox{.}(2005){Uytterhoeven}, {Harmanec},
  {Telting}, \& {Aerts}}]{uytterhoeven2005}
{Uytterhoeven} K., {Harmanec} P., {Telting} J.~H., {Aerts} C., 2005, \aap, 440,
  249

\bibitem[{{Vink} {et~al}\mbox{.}(2001){Vink}, {de Koter}, \&
  {Lamers}}]{vink2001}
{Vink} J.~S., {de Koter} A., {Lamers} H.~J.~G.~L.~M., 2001, \aap, 369, 574

\bibitem[{{Wade}(2015)}]{1411.3604}
{Wade} G.~A., 2015, in Proceedings of Magnetic Stars 2014, Special
  Astrophysical Observatory, pp. 373--374

\bibitem[{{Weber} \& {Davis}(1967)}]{wd1967}
{Weber} E.~J., {Davis}, Jr. L., 1967, \apj, 148, 217

\end{thebibliography}


\end{document}